\documentclass[sn-mathphys-num]{sn-jnl}

\usepackage{graphicx}%
\usepackage{subfig}
\usepackage{multirow}%
\usepackage{amsmath,amssymb,amsfonts}%
\usepackage{amsthm}%
\usepackage{siunitx}
\usepackage{mathrsfs}%
\usepackage[title]{appendix}%
\usepackage{xcolor}%
\usepackage{textcomp}%
\usepackage{manyfoot}%
\usepackage{booktabs}%
\usepackage{algorithm}%
\usepackage{algorithmicx}%
\usepackage{algpseudocode}%
\usepackage{listings}%
\usepackage{booktabs}
\usepackage{microtype}
\usepackage[version=3]{mhchem}

\usepackage[acronym]{glossaries}

\newacronym{ai}{AI}{Artificial Intelligence}
\newacronym{nn}{ANN}{Artificial Neural Network}
\newacronym{ts}{TS}{Thomson Scattering}
\newacronym{ece}{ECE}{Electron Cyclotron Emission}
\newacronym{dl}{DL}{Deep Learning}
\newacronym{co2}{\ce{Interferometer}}{\ce{CO2} interferometer}
\newacronym{mse}{MSE}{Motional Stark Effect}
\newacronym{cer}{CER}{Charge Exchange Recombination}
\newacronym{rcn}{RCN}{Reservoir Computing Network}
\newacronym{rnn}{RNN}{Recurrent Neural Network}
\newacronym{ae}{AE}{Alfv\'{e}n Eigenmode}
\newacronym{mlp}{MLP}{Multilayer Perceptron}
\newacronym{elm}{ELM}{Edge Localized Mode}
\newacronym{gpu}{GPU}{Graphics Processing Unit}
\newacronym{rmp}{RMP}{Resonant Magnetic Perturbation}
\newacronym{ml}{ML}{Machine Learning}
\newacronym{cnn}{CNN}{Convolutional Neural Network}
\newacronym{lstm}{LSTM}{Long-Short term memory cell}
\newacronym{mag}{Magnetics}{Magnetic probes}
\newacronym{fpp}{FPP}{Fusion Pilot Plant}
\newacronym{iter}{ITER}{International Thermonuclear Experimental Reactor}
\newacronym{srts}{SRTS}{Multimodal Super-Resolution TS}

\newcommand{\ddd}{DIII-D }

\theoremstyle{thmstyleone}%
\theoremstyle{thmstyletwo}%

\theoremstyle{thmstylethree}%

\begin{document}




\title {Multimodal Super-Resolution: Discovering hidden physics and its application to fusion plasmas}


\author*[1]{\fnm{Azarakhsh} \sur{Jalalvand}}\email{azarakhsh.jalalvand@princeton.edu}
\author[2]{\fnm{SangKyeun} \sur{Kim}}
\author[3]{\fnm{Jaemin} \sur{Seo}}
\author[2]{\fnm{Qiming} \sur{Hu}}
\author[1]{\fnm{Max} \sur{Curie}}
\author[1]{\fnm{Peter} \sur{Steiner}}
\author[4]{\fnm{Andrew Oakleigh} \sur{Nelson}}
\author[5]{\fnm{Yong-Su} \sur{Na}}
\author*[1,2]{\fnm{Egemen} \sur{Kolemen}}\email{ekolemen@princeton.edu}


\affil*[1]{\orgdiv{Department of Mechanical and Aerospace Engineering}, \orgname{Princeton University}, \orgaddress{\city{Princeton},\country{USA}}}

\affil[2]{\orgname{Princeton Plasma Physics Laboratory}, \orgaddress{\city{Princeton}, \country{USA}}}

\affil[3]{\orgdiv{Department of Physics}, \orgname{Chung-Ang University}, \orgaddress{\city{Seoul}, \country{South Korea}}}

\affil[4]{\orgdiv{Applied Physics and Applied Mathematics}, \orgname{Columbia University}, \orgaddress{\city{New York}, \country{USA}}}

\affil[5]{\orgdiv{Department of Nuclear Engineering}, \orgname{Seoul National University}, \orgaddress{\city{Seoul}, \country{Republic of Korea}}}

\abstract{
A non-linear complex system governed by multi-spatial and multi-temporal physics scales cannot be fully understood with a single diagnostic, as each provides only a partial view and much information is lost during data extraction. Combining multiple diagnostics may lead to incomplete projections of the system's physics. By identifying hidden inter-correlations between diagnostics, we can leverage mutual support to fill in these gaps, but uncovering these inter-correlations analytically is too complex. We introduce a groundbreaking machine learning methodology to address this issue. Unlike traditional methods, our multimodal approach does not rely on the target diagnostic's direct measurements to generate its super-resolution version. Instead, it utilizes other available diagnostics to produce super-resolution data, capturing detailed structural evolution and responses to perturbations that were previously unobservable. This capability not only enhances the resolution of a diagnostic for deeper insights but also reconstructs the target diagnostic, providing a valuable tool for mitigating diagnostic failure. This methodology addresses a critical problem in fusion plasmas: the \gls*{elm}, a plasma instability that can cause significant erosion of plasma-facing materials. One method to stabilize \gls*{elm} is using resonant magnetic perturbation to trigger magnetic islands. However, low spatial and temporal resolution of measurements limits the analysis of these magnetic islands due to their small size, rapid dynamics, and complex interactions within the plasma.
With super-resolution diagnostics, we can experimentally verify theoretical models of magnetic islands for the first time, providing unprecedented insights into their role in \gls*{elm} stabilization. This advancement aids in developing effective \gls*{elm} suppression strategies for future fusion reactors like ITER and has broader applications, potentially revolutionizing diagnostics in fields such as astronomy, astrophysics, and medical imaging.
}



\maketitle


\glsresetall
\section{Introduction}
\label{sec:Introduction}
In complex physical systems, diagnostic measurements are often intricately interconnected through fundamental physical principles. These underlying connections stem from the laws of nature that govern the behavior of matter and energy. For instance, electromagnetic events couple the measured signals, and equations of state relate variables such as pressure, volume, and temperature, providing a framework to infer one quantity from others. Similarly, coupled differential equations in fluid dynamics or plasma physics describe how multiple system parameters evolve interdependently over time. Such relationships are particularly evident in fusion energy, the focus field of this work, which is characterized by its intricate interplay of various physical phenomena.

The fusion energy technology aimed at producing eco-friendly energy is rapidly advancing through the synergy of academia and industry \cite{sciadv.abq5273,Degrave2022_deepmind, sciadv.aav2002,Waldrop2014_nature_fusionstartups}. The success of fusion energy is fundamentally based on maintaining high-temperature, high-pressure hydrogen plasma without becoming unstable. It was recently shown that \gls*{ai} can be a helpful tool to achieve that goal \cite{Seo2024avoiding, kim_highest_2024, PhysRevE.109.045201, Degrave2022_deepmind}. Fusion experimental facilities like \ddd \cite{Ding2024_d3dnature} utilize various diagnostics for effective plasma monitoring necessary for this \gls*{ai} application \cite{Boivin2005diii-d}. 
%
For example, the \gls*{ece} diagnostic system measures electron temperature \cite{Haskey2022details}, \gls*{co2} measures electron density and its fluctuations \cite{Strait2006magnetic}, \gls*{mse} measures the magnetic field \cite{Holcomb2006motional}, and \gls*{ts} measures the electron temperature and density \cite{Ponce-Marquez2010thomson}. The different measurements each capture different physical properties, and form a complementary set for extracting as much information from the plasma as possible. Although it is likely that there exists some kind of correlation or coupling between the measurements of different diagnostics (more details in Section Supplementary materials), our current scientific understanding is still not capable of specifying some of these relationships analytically. \gls*{ml} is a powerful tool for identifying hidden relationships in data \cite{Shousha_2024_ml}. Learning the hidden relationships among different diagnostics would be a great asset to enhance their measurements, and it also helps to find a minimal set of diagnostics for a future reactor in which the availability of diagnostics is limited due to the cost and hardware constraints.

One of the most critical issues for fusion reactors is the \gls*{elm}, an instability that occurs at the plasma edge under high-confinement conditions. This edge instability delivers transient and intense heat flux outward, which can cause unacceptable levels of erosion of plasma-facing materials in a reactor-scale device. Therefore, understanding and controlling this phenomenon is a major challenge that must be resolved \cite{kim_highest_2024, Joung_NF_2024}. However, the detailed physical mechanism of \gls*{elm}s and structure of the response to the external field occurring within milliseconds are still subjects of ongoing debate. High-frequency diagnostics like \gls*{ece} and \gls*{co2} possess sufficient time resolution to track these fast dynamics, but their limited spatial resolution and measurement conditions pose challenges in clearly observing the structural characteristics of \gls*{elm}s. On the other hand, \gls*{ts} offers high spatial resolution capable of observing detailed structures, but its temporal resolution is too low to elucidate the exact mechanism of \gls*{elm}s.

The current remedy to this issue is a specific operational method for \gls*{ts}, known as ``bunch mode'', to increase the sampling rate of  up to \SI{10}{\kHz} \cite{diallo2015,HE_2019}. Despite its high pulse repetition, firing \gls*{ts} in ``bunch mode'' is limited by the heat capacity of the laser medium and limited measurement repetition. Therefore such an approach is typically reserved for very short periods of time or specific experiments where high-resolution temporal data is crucial \cite{HE_2019}.

Instead, we hypothesize that a data-driven \gls*{ml} model, so-called \textbf{Diag2Diag}, with multimodal inputs comprising the high-frequency diagnostics can effectively make use of internal correlations in order to estimate \gls*{ts}. This can enhance the temporal resolution of the existing \gls*{ts} diagnostics without upgrading hardwares, so-called \textbf{\gls*{srts}} diagnostics, which enables deeper physical analysis of plasma behavior.

Various fields have developed \gls*{ml}-based spatial or temporal resolution enhancement techniques, but these mostly involve resolution enhancement by learning linear or nonlinear interpolation within single or limited types of data \cite{fasseaux_machine_2024,ward_machine_2022,pmlr-v145-balestriero22a,fukami_super-resolution_2023, REN2023112438, sciadv.adn0139}. These are applicable only to regularly sampled data and are challenging to generate finer-scale phenomena undetectable at the time resolution of the target sensor (more details in Section Supplementary materials). Our work goes beyond plausible interpolation; it is a physics-preserving super-resolution to reconstruct events missed by target diagnostics, by learning the correlation between different diagnostic measurements in fusion devices, which is, to our best knowledge, the first attempt of its kind.




\begin{figure}[!htb]
    \centering
    \includegraphics[width=\linewidth]{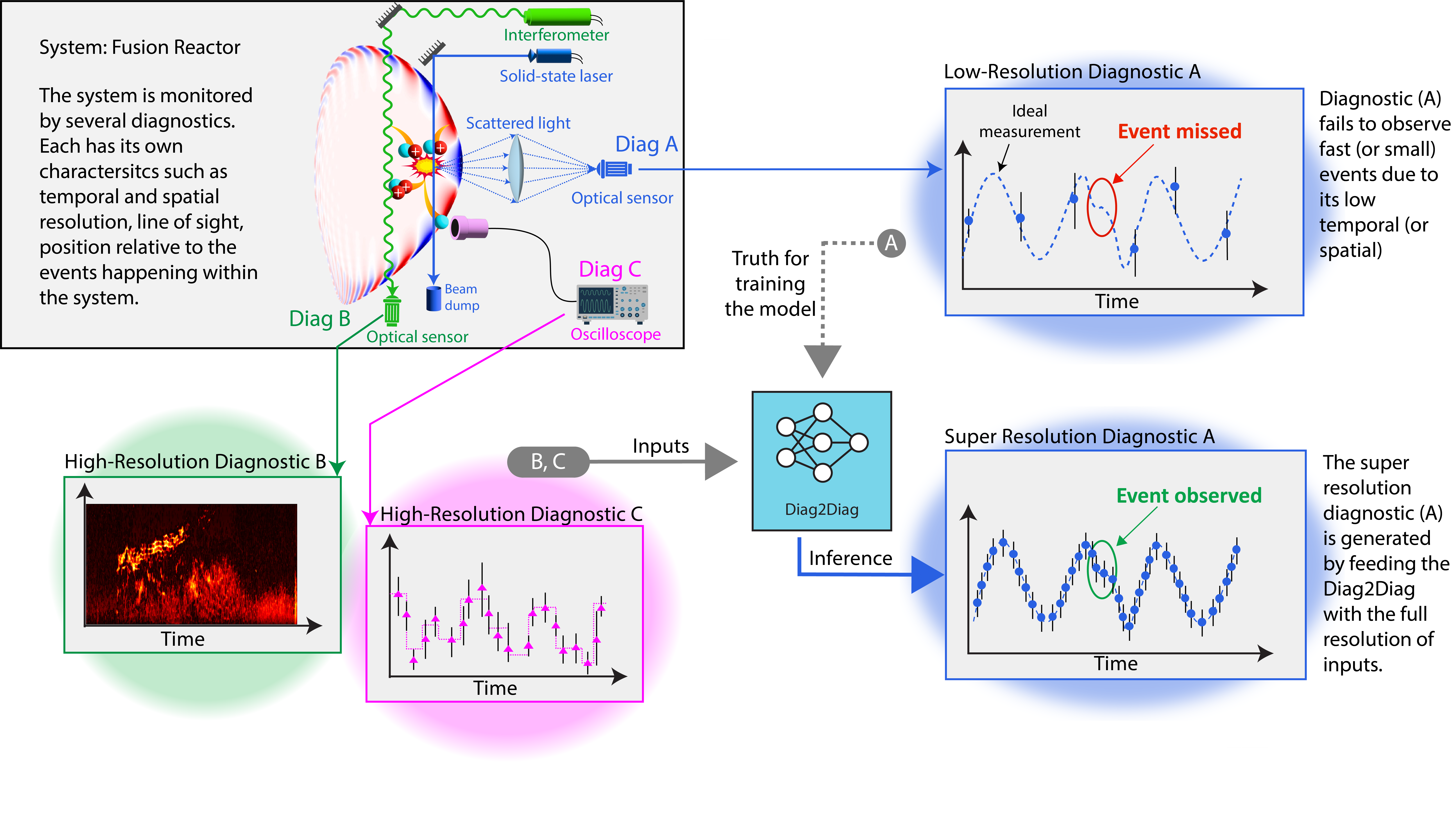}
    \caption{Main methodology. DiagA is essential to capture fast transient events near the edge of plasma. But due to its low temporal resolution and accuracy it fails to track the evolution of such events. Diag2Diag solves this problem by generating synthetic super resolution of DiagA by learning the correlation between DiagA data and other diagnostic measurements with higher resolutions and better accuracy.
    }
    \label{fig:MainMetholodogy}
\end{figure}


Figure \ref{fig:MainMetholodogy} summarizes the main methodology for this work. \ddd  utilizes hundreds of diagnostics for monitoring the plasma. These diagnostics measure various characteristics of plasma at different temporal resolutions. A potential \gls*{ml} model can learn the intrinsic correlations among diagnostics data and thus generate one from others. This works for both, time-series and spectrograms, although different variants of \gls*{nn} are used. The design choices and the optimization and training strategies are described in the following sections. 

\section{ML-based mapping between different diagnostics}
\label{sec:DiagnosticMapping}

For developing an \gls*{ml}-based \gls*{srts} diagnostic from other diagnostics, it is essential to verify the existence, strength, and robustness of correlation among them. We therefore approach this in several steps as described subsequently.

\begin{figure}[!htb]
    \centering
    \includegraphics[width=\linewidth]{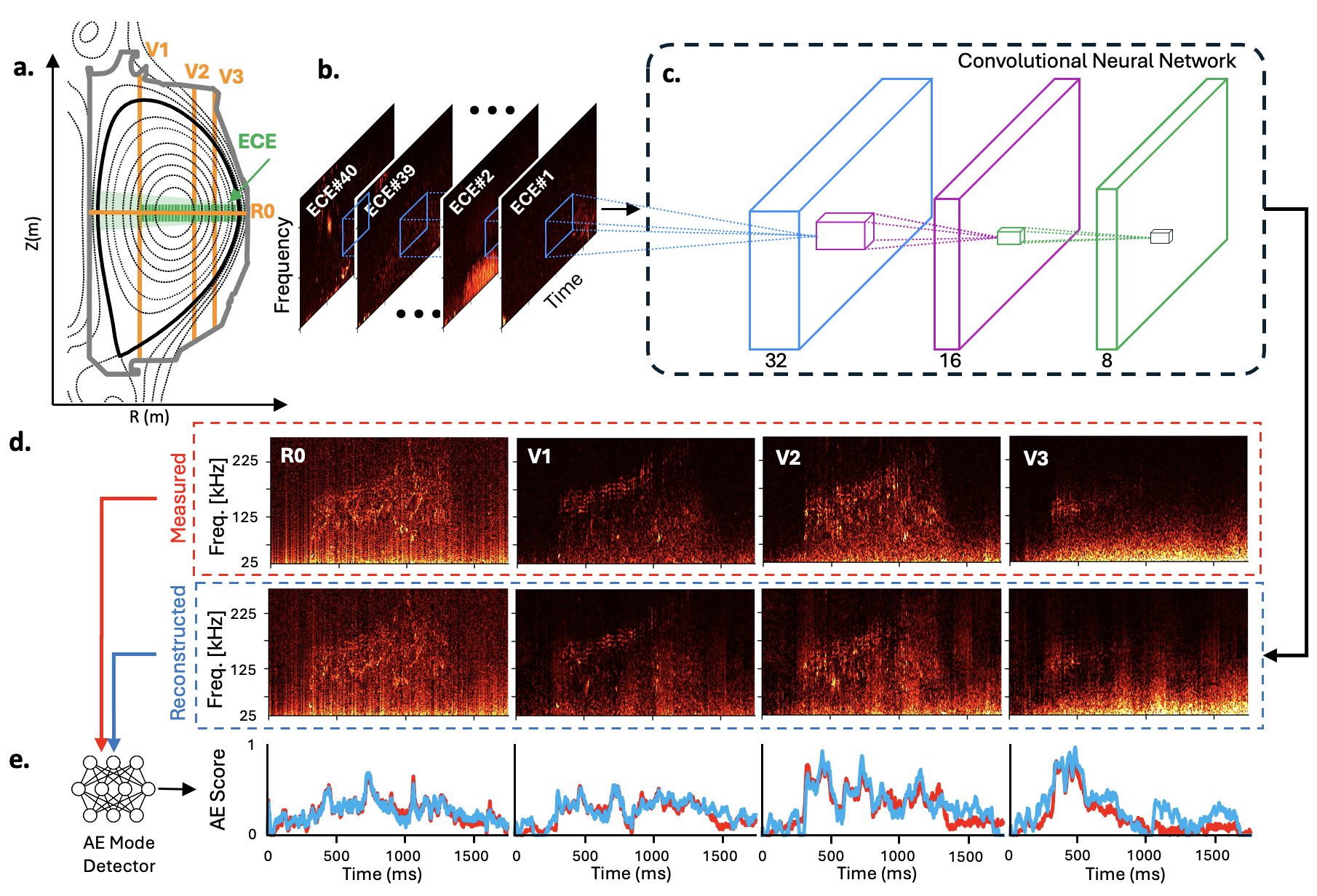}
    \caption{Reconstructing \gls*{co2} spectrograms from \gls*{ece} spectrograms of \ddd shot 170669 using convolutional neural networks. (a) The configuration of 4 \gls*{ece} and 40 \gls*{ece} probes at \ddd. (b) A tensor of $(40 \times time \times frequency)$ is supplied to CNN. (c) The configuration of CNN. (d) Visual comparison of measured and reconstructed spectrograms (e) Comparison of the Alfven Eigenmode detector output \cite{Garcia2023comparison} supplied with the measured and reconstructed spectrograms.}
    \label{fig:MainIdeaSpectrogramReconstruction}
\end{figure}

The aim of the first step is to show that we can reconstruct the spectrograms of one diagnostic based on another. As was discussed in the introduction, it is very likely that different diagnostics data have intrinsic correlations. Certain plasma instabilities and modes, such as \gls*{elm} and \gls*{ae}, affect both electron temperature, which is measured by \gls*{ece}, and density fluctuations measured by \gls*{co2}.
We now show that an \gls*{nn} is able to learn this relationship during \gls*{ae} modes by mapping from \gls*{ece} spectrograms to \gls*{co2} spectrograms as illustrated in Figure \ref{fig:MainIdeaSpectrogramReconstruction}. 

Figures \ref{fig:MainIdeaSpectrogramReconstruction}(b) and (d) show example spectrograms obtained from the raw signals of \gls*{ece} and \gls*{co2}, respectively, whose measurement positions and paths can be seen in Figure \ref{fig:MainIdeaSpectrogramReconstruction}(a). We designed and trained a \gls*{cnn} that takes 40 \gls*{ece} spectrograms as input and reconstructs 4 target \gls*{co2} spectrograms, as shown in Figure \ref{fig:MainIdeaSpectrogramReconstruction}(c). The reconstructed synthetic \gls*{co2} spectrograms visually confirm the plausible reconstruction of features such as frequency chirping and harmonics as seen in Figure \ref{fig:MainIdeaSpectrogramReconstruction}(d).

Besides the visual comparison, we are also interested in how much the underlying physical information is preserved using this method. Therefore, we evaluated the preservation of physical information by performing a downstream task, \gls*{ae} instability detection \cite{Garcia2023comparison}, based on the measured and the reconstructed \gls*{co2} spectrograms. In Figure \ref{fig:MainIdeaSpectrogramReconstruction}(e), we can see that the \gls*{ae} scores obtained from the reconstructed spectrograms (blue) closely match those from the measured spectrograms (red). This demonstrates that the results generated by \gls*{ml} contain sufficient hidden physical information, and thus it is supported that \gls*{ml} can extract the intrinsic correlation among diagnostic data. After this initial study with visualization on the spectrogram domain, we now shift to time-series domain and a new task to tackle the generation of \gls*{ts} signals from other diagnostics based on raw time-series, not a spectrogram.



\section{Multimodal super-resolution diagnostic}
In this section, we switch from spectrograms to time-series signals and show that the amplitude of a diagnostics can be reconstructed from other diagnostics, while preserving intrinsic physics. More importantly, we will show that if the input diagnostics are of much higher temporal resolution compared to the target one, such a model can be used to increase the time resolution of the target signals in a much more intelligent way compared to the conventional uni-modal interpolations. As a use case, we target \gls*{ts}, one of the most important diagnostics that measure the electron density and electron temperature profile of plasma. However as mentioned earlier, its low temporal resolution is a bottleneck in studying the plasma evolution in the rapidly changing events such as \gls*{elm}.

We consider a suite of input diagnostics available at \ddd including \gls*{co2}, \gls*{ece}, \gls*{mag}, \gls*{cer}, and \gls*{mse} with typical sampling rates of \SI{1.66}{\MHz}, \SI{500}{\kHz}, \SI{2}{\MHz}, \SI{200}{\Hz}, and \SI{4}{\kHz}, respectively. Since our aim is not only to enhance but also to reconstruct \gls*{ts} from other diagnostics, we do not use the available measurement of this diagnostic as input to Diag2Diag. To obtain a dataset suitable for this task, all the included diagnostics are aligned with the \gls*{ts} sampling time steps by matching their most recent measured sample. In this way, we create a dataset with which we train the Diag2Diag \gls*{nn} for this task. 
Since the sampling steps of \gls*{ts} are not always uniform in time (See Figure \ref{fig:slow_ts}), we opted for a memory-less neural network instead of the recurrent neural network commonly used in time-series analysis. However, we included the first and second derivatives of the high-resolution input diagnostics, \gls*{ece} and \gls*{co2}, to include the temporal evolution information. 



\begin{figure}[!htb]
    \centering
    \includegraphics[width=.8\linewidth]{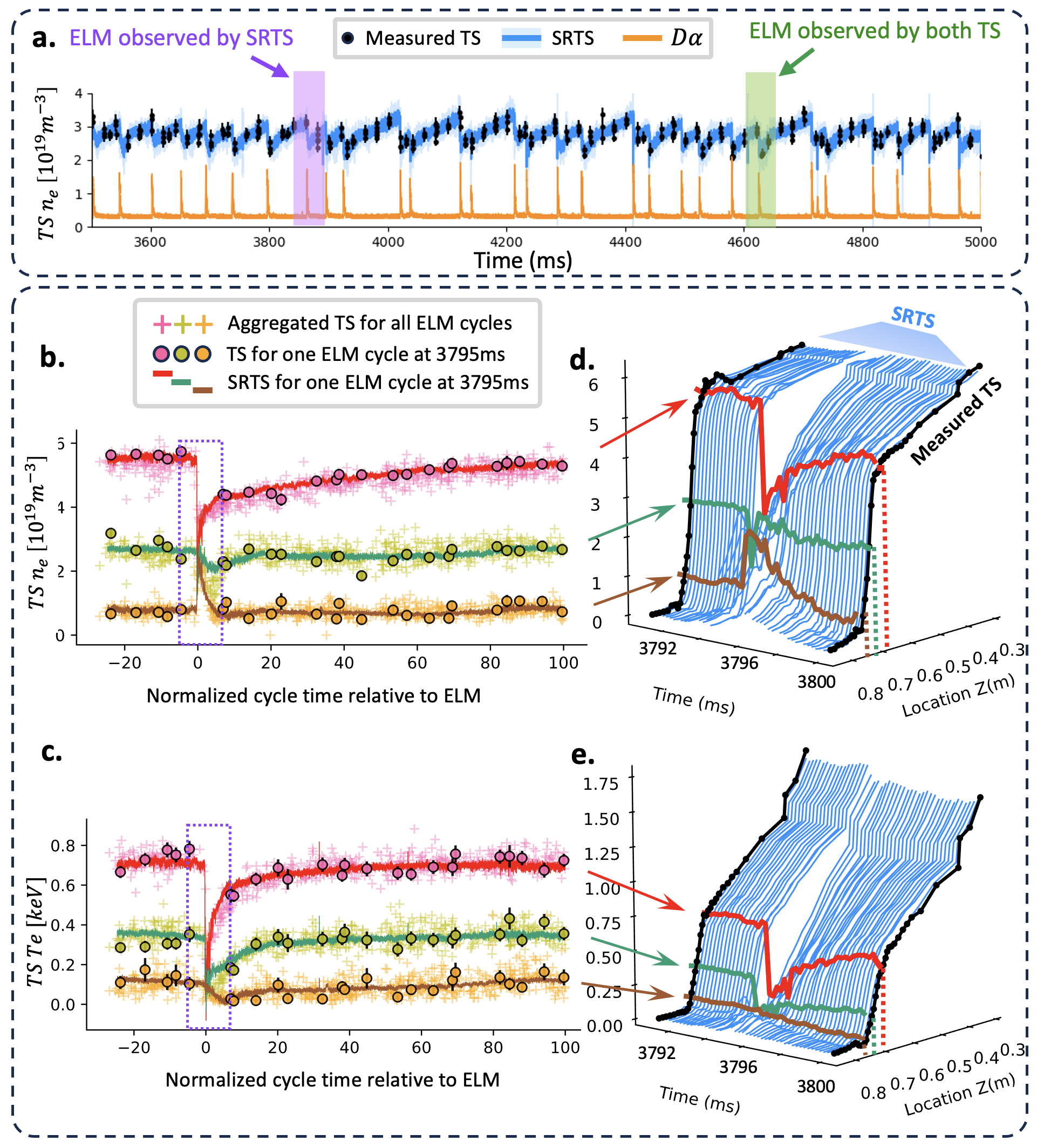}
    \caption{(a) Comparison of the electron density by the measured \gls*{ts} and the synthetic \gls*{srts}, for the \ddd shot 153761 \cite{diallo2015} near the edge ($Z=0.71m$). $D_\alpha$ with arbitrary units is plotted as an indicator of \gls*{elm}s. An example of \gls*{elm} event captured by both diagnostics, and another example only captured by \gls*{srts} are highlighted in green and purple, respectively. (b-c) Aggregating the measured \gls*{ts} density and temperature in three locations of plasma near the edge for several \gls*{elm} cycles of the \ddd shot 174832. The circle highlights the measures \gls*{ts} for one selected \gls*{elm} cycle and the solid lines present the \gls*{srts} which agreeably match the measures \gls*{ts}. $t=0$ represents the time when \gls*{elm} is identified by $D_\alpha$. (d-e) The evolution of \gls*{srts} between two consecutive measured \gls*{ts} near one \gls*{elm} cycle across the plasma location.}
    \label{fig:TS_hi_TS_Z}
\end{figure}

Figure \ref{fig:TS_hi_TS_Z}(a) shows, in blue, synthetic \gls*{ts} signals (or \gls*{srts}) reconstructed through the trained Diag2Diag from other high-frequency diagnostics, where the original \gls*{ts} signals are also shown with black dots. We can observe that the synthetic signals closely follow the original signals. Diag2Diag’s ability to reconstruct \gls*{ts} from other diagnostics ensures that crucial information is not lost, even in the absence of direct measurements. Furthermore, while the original signals sometimes fail to capture \gls*{elm} events (identifiable by spectral emission ($D_{\alpha}$)), the synthetic signals accurately capture the events missed between the original signals.

\subsection*{Validation: Investigating ELM cycles in \ddd}

When \gls*{elm} instability occurs, a large amount of plasma quickly escapes from the boundary within milliseconds, and then the plasma gradually recovers. \gls*{ts} diagnostics can observe the density and temperature structure at this edge region, but are limited in capturing dynamics occurring over milliseconds. Recent research \cite{Nelson2021interpretative} overcame these resolution limits by aggregating the measurements from multiple repeated cycles of the fast activity under almost identical conditions to observe a complete evolution. In this subsection, over 20 highly reproduced cycles of \gls*{elm} crash and recovery were aggregated from \ddd shot 174823 to assume a ground truth of a complete evolution of an \gls*{elm} cycle.

The aggregated density and temperature evolution measured by \gls*{ts} in three locations of plasma near the edge are shown in Figures \ref{fig:TS_hi_TS_Z}(b-c) with transparent crosses, while measurements from a single cycle are shown as solid dots, with different colors for different measurement locations.

We used the Diag2Diag model to generate synthetic \gls*{srts}, shown with solid lines in Figure \ref{fig:TS_hi_TS_Z}(b-c). The \gls*{srts} signal from a single cycle around time \SI{3795}{ms} not only follows the trend of the aggregated multiple \gls*{ts} measurements but also well overlays the \gls*{ts} measurements within that cycle. Figures \ref{fig:TS_hi_TS_Z}(d-e) show the detailed evolution of plasma density and temperature across the plasma plasma location captured by \gls*{srts} in the same ELM cycle at \SI{3795}{ms} which is missed by \gls*{ts} between its two consecutive measurements at \SI{3791}{ms} and \SI{3800}{ms}. 
 

In a more typical tokamak discharge, the plasma state continually changes, and \gls*{elm}s occur more irregularly, as shown in Figure \ref{fig:TS_hi_TS_Z}(a). In such cases, it is not possible to reconstruct a single \gls*{elm} cycle by aggregating multiple cycles, and our \gls*{srts} method will be highly beneficial.

\section{Science discovery: Unveiling diagnostic evidence of RMP mechanism on the plasma boundary} \label{sec:discovery}
In what follows, we investigate whether the synthetic super-resolution diagnostics can help to verify the hypotheses on the mechanism of plasma response to external field perturbations in fusion plasma physics that have been proposed theoretically or by simulations but have never been visualized with the experimental data due to the lack of diagnostic resolution. 

\begin{figure}[!htb]
    \centering
    \includegraphics[width=\linewidth]{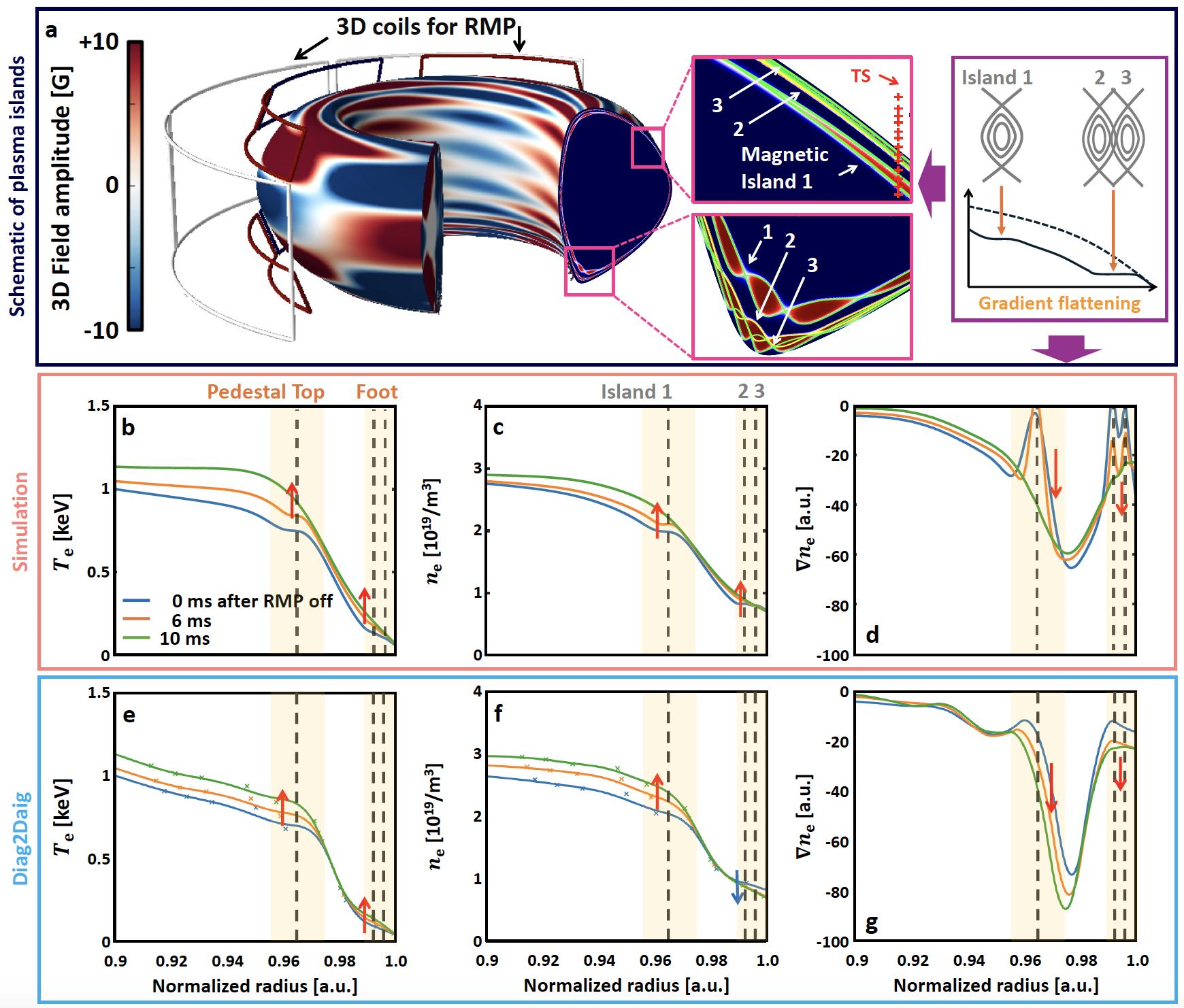}
    \caption{Structure of 3D coils and islands by perturbed field (a), and the evidence in the simulation (b-d) and the \gls*{srts} diagnostic (e-g) for \gls*{rmp}-induced island mechanism on the plasma boundary in \ddd shot 157545.} 
    \label{fig:TS_foot}
\end{figure}

One promising strategy to control \gls*{elm}s is employing \gls*{rmp}s \cite{JET-RMP-2007,MAST-RMP-2010,AUG-RMP-2011,EAST-RMP-2016,KSTAR-RMP-2018} generated by external 3D field coils depicted in Figure \ref{fig:TS_foot}(a). These fields effectively reduce the temperature and density at the confinement pedestal, stabilizing the energy bursts in the edge region. Consequently, ITER will rely on \gls*{rmp}s to maintain a burst-free burning plasma in a tokamak, making it essential for the fusion community to understand and predict its physics mechanism\cite{Loarte_2014}. However, this issue has remained a challenge for decades.

The leading theory \cite{Snyder-2012,nazikian_2016,Orain-2019,qiminghu2019} for explaining the reduced pedestal by \gls*{rmp}s is the formation of magnetic islands by an external 3D field. The magnetic island is a ubiquitous feature in an electromagnetic system with plasmas \cite{Furth-1963} formed by field reconnection \cite{Parker-1957,Sweet_1958}. This structure allows rapid heat (or temperature) and particle (or density) transport between adjacent magnetic field lines, strongly reducing the gradient of local heat and particle distribution or, in other words, profile flattening\cite{Fitzpatrick-1995}. The existing theories explain that RMP forms static magnetic islands at the pedestal top and foot region, therefore reducing the pedestal by local profile flattening. As illustrated in Figure \ref{fig:TS_foot}(a), the theory predicts that \gls*{rmp}s can create magnetic islands near the plasma boundary where the pedestal sits. This model has been successful in quantitatively explaining and predicting the RMP-induced pedestal degradation in real experiments \cite{ Orain_2019, Markl_2023}, reinforcing magnetic islands as a promising mechanism for \gls*{rmp}-induced pedestal degradation. Nevertheless, measuring evidence of island or local profile flattening still remains a challenge. Extensive experimental efforts have been conducted for this reason and were able to capture the local flattening electron temperature profile \cite{PhysRevLett.125.045001} near the pedestal top, strongly supporting this theory. However, simultaneously measuring electron temperature and density both at the pedestal top and foot was not possible. In a previous study, rough evidence was observed in \gls*{ts} \cite{qiminghu2019}, but it was insufficient to derive a concrete conclusion, mainly due to a large uncertainty of measurement originating from narrow structure (expected from theory, see Fig.\ref{fig:TS_foot}(a)) and oscillatory nature of the plasma boundary. To address the diagnostic uncertainties caused by such system oscillation, one method is to increase the time sampling rate and use time averaging. However, in conventional \gls*{ts}, increasing the time resolution results in a trade-off with measured accuracy, eventually leading to observational limitations.

Interestingly, the \gls*{srts} has once again illuminated the profile evolution by \gls*{rmp} application, providing the novel evidence of ''simultaneous'' flattening of temperature and density profile at both the top and the foot of the pedestal, strongly supporting the theoretical prediction of magnetic islands effect. This is possible by capturing the statically reliable time trace of the profile with the Chebyshev time filter, leveraging the enhanced temporal resolution by \gls*{srts}.

Figures \ref{fig:TS_foot}(b-g) illustrate the recovery of temperature and density pedestals within \SI{10}{\ms} after deactivating \gls*{rmp}, as captured through numerical modeling (Figure \ref{fig:TS_foot}(b-d)) and \gls*{srts} (Figure \ref{fig:TS_foot}(e-g)). The simulations reveal that the recovery of temperature and density pedestals begins at the top and foot, coinciding with the disappearance of islands. As depicted in Figures \ref{fig:TS_foot}(d) and (g), the profile gradient recovers at these island locations, enhancing the overall profile. For instance, the measured temperature pedestal shows recovery at both the top and foot through an increasing gradient, displaying qualitative alignment with the simulation results. However, some discrepancies are noted, particularly in the density evolution at the pedestal foot in the \gls*{srts}, even though its gradient remains consistent with the modeling. These quantitative differences may stem from the \gls*{ts}'s limited spatial resolution at the boundary and the modeling assumptions such as fixed boundary conditions \cite{Yu_2020}. Nevertheless, the gradient evolution directly indicates a change in transport due to the \gls*{rmp}-induced islands during this perturbative profile evolution, highlighting that the \gls*{srts} successfully reveals the experimental island effect. This provides the new diagnostic evidence of profile flattening at magnetic islands, a key mechanism of \gls*{rmp}-induced pedestal degradation. 

\begin{figure}[!htb]
    \centering
    \includegraphics[width=\linewidth]{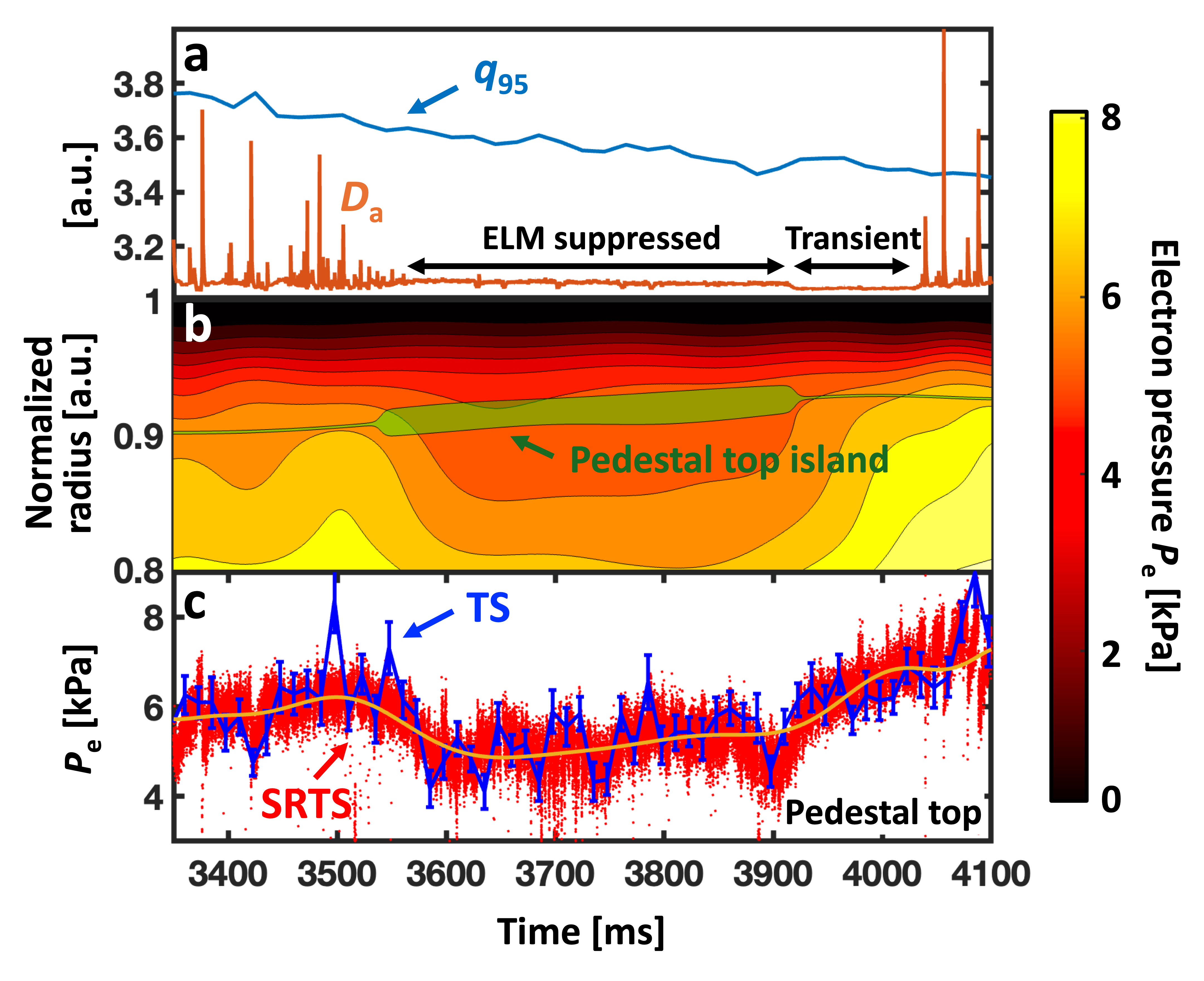}
    \caption{(a) Time evolution of edge safety factor ($q_\text{95}$)  and $D_\alpha$ emission at plasma edge. (b) Contour of electron pressure versus normalized plasma radius and time. The numerically derived width of the magnetic island at the pedestal top is illustrated as green contours. (c) Comparison of TS (blue), \gls*{srts} (red), and filtered \gls*{srts} (orange solid line).} 
    \label{fig:TS_q95}
\end{figure}

The strength of the \gls*{srts} in unveiling profile flattening during ELM suppression can be further highlighted with additional cases. Figure \ref{fig:TS_q95} shows the time traces of plasma when the edge safety factor ($q_\text{95}$), the magnetic pitch angle at the plasma edge, gradually decreases. Here, all other plasma operation parameters, including the RMP field, remain the same. From $D_\alpha$ emission looking perturbation of plasma edge (see \ref{fig:TS_q95}a), the bursty spikes disappear during $q_\text{95}=$3.5-3.6, corresponding to the ELM-suppressed phase followed by the transient ELM-free phase. This shows the strong dependence of ELM suppression on $q_\text{95}$. The modeling work based on the island physics \cite{PhysRevLett.125.045001} was able to explain this behavior through a sensitivity of island width at the pedestal top, where its width abruptly increases at certain $q_\text{95}$ values due to nonlinear RMP response \cite{10.1063-5.0043018,Hu_2020}. When the island becomes bigger, it leads to local flattening of electron pressure ($P_\text{e}$, product of temperature and density), resulting in ELM suppression. This explanation has successfully predicted this $q_\text{95}$ dependency in multiple devices \cite{10.1063-5.0043018}. However, its experimental validation remains challenging as plasma becomes perturbative while $q_\text{95}$ changes, making pedestal diagnostic oscillatory. Such diagnostic oscillation can be overcome by time filtering, but the temporal resolution of TS was limited for resolving pedestal evolution with $q_\text{95}$ with filtering processing.

The \gls*{srts} has once again derived the profile evolution by $q_\text{95}$ change, providing novel evidence of profile flattening of pressure profile at the top of the pedestal, leveraging the enhanced temporal resolution by \gls*{srts}. Figures \ref{fig:TS_q95}b illustrate the strong flattening of the pressure profile during the ELM-suppressed phase, coinciding with the location and width of the magnetic island from numerical modeling. Figures \ref{fig:TS_q95}c shows the electron pedestal height measured in both TS and \gls*{srts}, where the filtered \gls*{srts} (orange solid line) follows TS while overcoming diagnostic oscillations, successfully extracting the main behavior of the pedestal. This successful application of \gls*{srts} underscores its potential to reveal new physics beyond the limitations of conventional diagnostic techniques.

\section{Conclusion}
\label{sec:Discussion}

This study introduces a transformative approach in the field of signal processing and diagnostics through the development of a multimodal neural network, Diag2Diag, which significantly enhances temporal resolution. By leveraging the intrinsic correlations among various diagnostic measurements, we have demonstrated the potential to increase the temporal resolution of the Thomson Scattering diagnostics in fusion plasma from a standard \SI{0.2}{\kHz} to an unprecedented \SI{1}{\MHz}. This improvement has unlocked new potentials in analyzing fast transient phenomena in plasma, such as the \gls*{elm}s and the effects of \gls*{rmp}s on pedestal degradation, which were previously blurred or missed in lower resolution data. The ability to inspect these dynamics in greater detail provides new insights into plasma behavior, particularly in conditions where key physics is hidden in the milliseconds. This enhancement is not merely a technical improvement but a crucial enabler for deeper insights into plasma behaviors that are pivotal for advancing fusion reactors. Furthermore, the model’s ability to reconstruct and predict diagnostics from other available diagnostics opens new avenues for measurement failure mitigation, cost-effective and less hardware-dependent diagnostic systems. This is particularly beneficial for experimental setups where space and resources are limited, such as in smaller fusion test facilities or in environments where installing multiple high-resolution diagnostics is impractical.

The implications of this work extend well beyond the immediate application to magnetic fusion. The multimodal super-resolution capabilities developed here can significantly impact areas such as laser fusion data analysis, accelerator data analysis, and molecular dynamics research. In these fields, similar challenges exist where the time resolution of diagnostics is inadequate to capture fast phenomena effectively. By applying our method, researchers can potentially uncover new physical phenomena or confirm theoretical predictions that were previously unverifiable through experiments due to resolution constraints.

In conclusion, the Diag2Diag model not only addresses a critical need within the fusion community but also sets a precedent for the broader application of AI and machine learning in physical sciences. By pushing the boundaries of what can be observed and measured, this work contributes to the foundational technologies necessary for the realization of fusion energy and advances our understanding of complex physical systems across various scientific domains.
\section{Methods}
\label{sec:Methods}



\subsection{Underlying Physics in Coupling of Diagnostic Measurements in Plasma System}    
\label{subsec:diag_connections}
\begin{figure}[!htb]
    \centering
    \includegraphics[width=\linewidth]{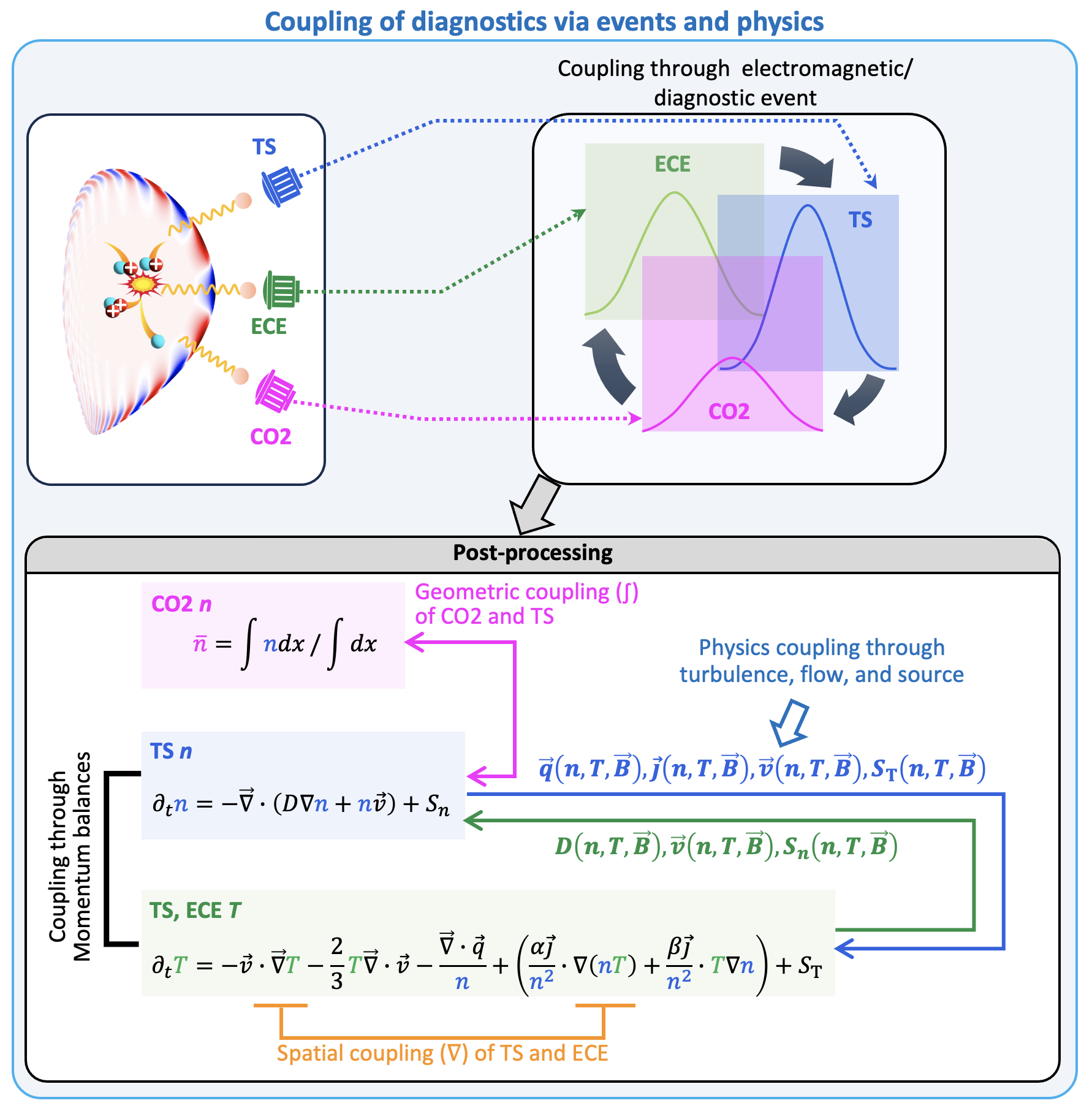}
    \caption{Schematic of couplings between diagnostics of the plasma system. They are connected through electromagnetic interactions between signals. Simultaneously, derived quantities from these signals are coupled via geometric definitions, momentum balances, and high-order physics, including turbulence, flow, and source in the system.} 
    \label{fig:dig_coupling}
\end{figure}

Diagnostics of electromagnetic systems involve measuring photons or waves to determine the physical quantities of these systems through post-processing. Due to the nature of the systems, these diagnostics are connected. Firstly, the measured signals are interconnected through electromagnetic interactions during system events. Additionally, the physical quantities obtained from signal processing are closely linked through momentum balances. Electromagnetic plasma quantities are governed by a series of momentum equations that encompass variables such as density, flow, temperature, and higher-order terms. Figure \ref{fig:dig_coupling} illustrates the momentum equations for plasma density ($n$) and temperature ($T$), where $D$ represents particle diffusivity, $v$ is plasma flow, $S_n$ is the particle source, $q$ denotes heat flux, $B$ stands for magnetic field, $j$ is plasma current, $S_T$ is the heat source, and ($\alpha$, $\beta$) are constant coefficients determined by plasma properties \cite{hutchinson2001introduction}. These equations demonstrate how the measured plasma quantities are interrelated both spatially and temporally. For instance, the line-averaged density obtained from \gls*{co2} diagnostics is geometrically linked to the local density measured by \gls*{ts} by its definition. Simultaneously, temperatures measured by \gls*{ts} and \gls*{ece} diagnostics, which are positioned differently, are spatially coupled through the gradient term in the momentum equations. Although the \gls*{ts} density and temperature do not directly interact in the equations, they are tightly linked via diffusive fluxes influenced by turbulence, flow, and sources in a self-consistent manner. This intricate physical coupling of various diagnostic measurements allows \gls*{ml} to identify and predict their interconnections effectively.

\subsection{Related works}    
\label{subsec:related}

In recent years different kinds of \gls*{nn} have been used for upsampling visual data \cite{wang2024physicsinformed,Jiang2018super,Niklaus2018context,Bao2019depth,Xue2019video} and for radar data \cite{Che2018hierarchical,Recla2022deep,Recla2023improving,Rossberg2023temporal}. These approaches are typically some kind of non-linear interpolation to add frames between existing video frames. More examples for \gls*{ml}-based upsampling were proposed for medical data \cite{Bellos2019convolutional} and for audio data \cite{Pascual2017segan,Donahue2018adversarial,Pandey2020densely,Kumar2019melgan}. Similar to the video upsampling approaches, these approaches can be considered a subcategory of non-linear interpolation as well. In \cite{Yoon2019estimating}, an alternative to interpolation is suggested to estimate missing data in temporal data streams. It is to some extent a multimodal approach, because it fuses different kinds of information. However, the algorithm is limited towards estimating missing data or dealing with irregularly sampled data. Approaches like these work well for enhancing existing sequences, which are quasi-stationary in a way such that consecutive frames or samples do not change very fast. 

However, in fusion energy, many spurious events like \gls*{elm} can happen between two \gls*{ts} samples. By interpolating between consecutive \gls*{ts} samples, regardless of linearly or non-linearly, it is likely that we would miss such spurious events. In our work, we thus develop a novel method to generate additional \gls*{ts} samples based on other diagnostics. This is roughly inspired by other multimodal \gls*{ml} approaches, such as \cite{Li2024farfusion}, where it was proposed to fuse Radar and camera data for an enhanced distance estimation. This is a multimodal approach and thus related to our approach, or \cite{Melis2024machine}, where machine learning was used to reveal the control mechanics of an insect wing hinge. This was also a multimodal approach in a way that the \gls*{ml} algorithm received different features recorded from flying insects. However, similar to the other approaches, no attempts to upsampling or estimating missing/in-between data are made. Also \cite{Player_2022} presents an artificial neural network method to enhance historical electron temperature data from the decommissioned C-2U fusion device. The model significantly increases the effective sampling rate of \gls*{ts} temperature measurements, utilizing data from multiple diagnostics including the measured TS. The method's effectiveness is demonstrated through comparisons with ensemble-averaged data for micro-burst instability study. The model's main drawbacks include limited generalization to only temperature profile study for one specific plasma regime. Notably, the work does not explore the model's potential for discovering new physics in fusion plasmas.

\subsection{Diagnostic details for ELM}    
\label{subsec:ELM_diag}

In order to let fusion energy be a viable energy source, it must achieve significant fusion gain through continuous fusion reactions. A prominent method to reach this objective is operating a tokamak in high-confinement mode (H-mode), which has a narrow edge transport barrier, also known as the pedestal. This feature significantly boosts plasma confinement within the reactor, enhancing fusion power and efficiency. However, operating in H-mode introduces a steep pressure gradient at the pedestal, leading to substantial operational risks. This gradient drives hazardous edge energy bursts due to a plasma instability known as \gls*{elm}s. These bursts lead to sudden drops in the energy at the pedestal, causing severe, transient heat fluxes on the reactor walls. This results in damaging material, potential surface erosion and melting, with heat energy reaching approximately \SI{20}{\mega\joule\per\m\squared}, which is an unacceptable level for fusion reactors. From ITER, future machines will not allow even the first \gls*{elm}. Therefore, to advance tokamak designs toward practical application in fusion energy, it is crucial to develop dependable methods to consistently suppress these edge burst events.

\begin{figure}[!htb]
    \centering
    \includegraphics[width=\linewidth]{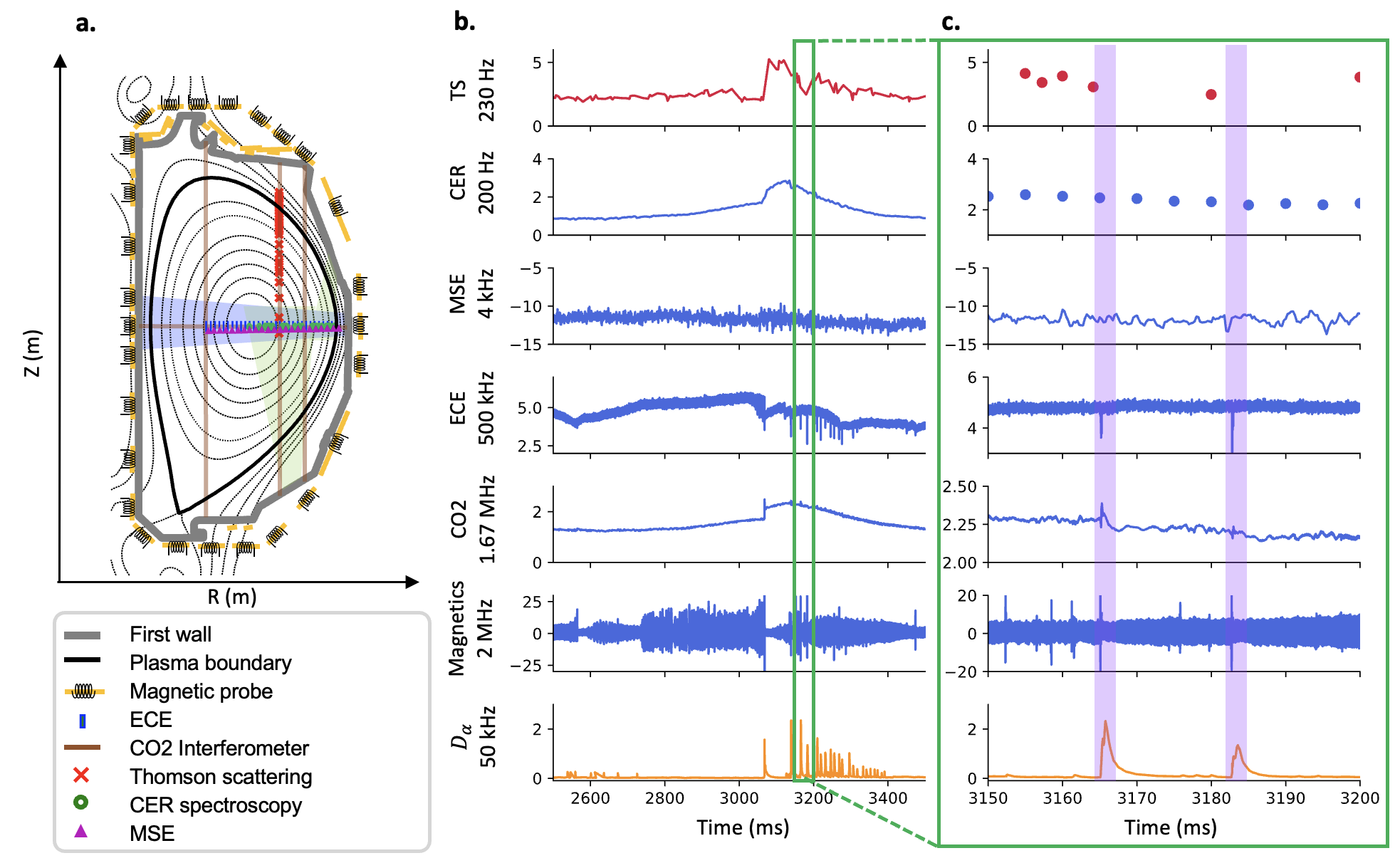}
    \caption{(a) Configuration of some diagnostics at \ddd. (b) Example of \gls*{ts} signal for \ddd discharge 190736, in red, along with the $D_\alpha$ measurements, as an indicator of \gls*{elm}, and a collection of other diagnostics that is used to increase the resolution of \gls*{ts}. (c) The same data as (b) but zoomed in to show the sampling points of the diagnostics around two examples of \gls*{elm} event. Due to the low sampling rate, this \gls*{elm} is not observed by \gls*{ts}. However, thanks to their high temporal resolution,  diagnostics including \gls*{ece}, \gls*{co2}, and \gls*{mag} capture that.} 
    \label{fig:slow_ts}
\end{figure}

A limitation of some diagnostics, such as \gls*{ts} is the low temporal resolution of only \SI{200}{\Hz}, which does not allow for detecting and tracking fast events like \gls*{elm} ($\leq \SI{1}{\ms}$). Figure \ref{fig:slow_ts} shows an example of missing \gls*{elm} in a  discharge, due to the low temporal resolution of \gls*{ts}. Nevertheless, it is still important to detect such events reliably, as they can have a strong impact on plasma behavior. 


On the other hand, diagnostics like \gls*{co2}, \gls*{ece}, have much higher temporal resolution with sampling frequencies around \si{\MHz}, which allows for a much more detailed analysis of the plasma. However, these diagnostics have different characteristics compared to \gls*{ts}. While \gls*{ts} offers detailed insights into both electron density and temperature with high accuracy, it requires complex setups and is usually more resource-intensive. A \gls*{co2} provides a more straightforward approach to measuring electron density, excelling in situations that require rapid response and continuous monitoring. Furthermore, \gls*{ece} and \gls*{ts} are both pivotal diagnostic tools used in tokamaks for measuring electron temperature, yet they operate on distinctly different principles and offer unique advantages. \gls*{ece} utilizes the natural microwave emissions from electrons gyrating around magnetic field lines to provide excellent temporal resolution, allowing for the monitoring of rapid plasma changes and instabilities, though its effectiveness can be limited by variations in magnetic field strength. On the other hand, \gls*{ts} involves firing a laser into the plasma and analyzing the scattered light, which provides robust, absolute measurements of both electron temperature and density with less susceptibility to magnetic influences. While \gls*{ece} excels in continuous data collection and fine temporal analysis, \gls*{ts} offers superior spatial resolution and is less dependent on external conditions, making it invaluable for comprehensive, though typically less frequent, plasma evaluations. If it would be possible to find a correlation between the measurements of those high-resolution diagnostics and \gls*{ts}, this would be useful for developing new physical analyses.

\subsection{Data acquisition}    
\label{subsec:DataAcquisition}

For this experiment, we used discharges from the \ddd tokamak that include all data from the key diagnostics of interest (\gls*{cer}, \gls*{co2}, \gls*{ece}, \gls*{mse}, and \gls*{ts}). We randomly selected \num{4000} discharges recorded between the years \num{2017} and \num{2022} to ensure a diverse and representative dataset. The diagnostic data was collected using the \ddd MDSplus \cite{Fredian2002mdsplus} and PTDATA \cite{Schissel2000data} systems. These diagnostics are generally provided as time-series data streams with varying sampling frequencies, ranging from \SI{200}{\Hz} for \gls*{ts} up to \SI{1.66}{\MHz} for \gls*{co2}.  The specific pre-processing steps applied to the data for the different experiments conducted in this study are detailed in the following sections.

\subsection{Feature extraction}
\label{subsec:FeatureExtraction}

For the spectrogram experiments, we consider the \gls*{co2} and \gls*{ece} diagnostics. We compute logarithmic magnitude spectrogram from time-series of the raw diagnostics. For each channel (\num{40} \gls*{ece} channels and \num{4} \gls*{co2} channels), we therefore used hamming windows of \SI{1}{\ms} with \SI{0.5}{\ms} overlap. In this way, it was ensured that the different magnitude spectrogrames are aligned in time. The spectrograms were afterwards converted to a logarithmic scale, clipped and rescaled to the range of [0, 1]. Given the noisy nature of the \gls*{ece} signals and after rescaling the spectrograms to the range of [0,1], the spectrograms are enhanced using a pipeline of image processing filters that includes
\begin{itemize}
    \item Quantile Filtering with a threshold of 0.9,
    \item Gaussian Blur Filtering on patches of size 31x3,
    \item Subtracting average per frequency bin
\end{itemize}   



We used the \gls*{ece} spectrograms as inputs to our model. Since we treated every \gls*{ece} channel independently during feature extraction, we obtained one spectrogram per channel, resulting in \num{40} input spectrograms (one per \gls*{ece} channel). Since our model is designed to estimate the \gls*{co2} spectrograms, it predicts four output spectrogram channels corresponding to the four \gls*{co2} interferometer channels.

For the time-series models, the different diagnostic measurements have varying sampling rates, and some are even non-uniformly sampled in time. Since the aim of time-series data analysis was to increase the resolution of \gls*{ts}, we used its timestamps as a reference and aligned all diagnostic modalities to \gls*{ts} by matching their most recent measured samples in time. This resulted in an amount of \num{135233} training, \num{22084} validation, and \num{18721} test samples.

For \gls*{co2} and \gls*{ece}, we also included the first and second temporal derivatives. Therefore, we smoothed the signals with a moving average window of \SI{1}{\ms} (\num{1660} \gls*{co2} samples and \num{500} \gls*{ece} samples), and then computed the first and second temporal derivatives of the smoothed signal also with a window of \SI{1}{\ms}. In this way, we can consider a temporal context of \SI{4}{\ms}.

The diagnostics \gls*{cer} and \gls*{mse} have a low temporal resolution, i.e., sampling frequencies of \SI{200}{\Hz} and \SI{4}{\kHz}, respectively. In this paper, we assume that they evolve only slowly in time. For the upsampling experiments, we thus pad these diagnostics after a measured sample with constant values until the next measured sample arrived.

The diagnostics (\gls*{cer}, \gls*{co2}, \gls*{ece}, and \gls*{mse}) together with the derivatives of \gls*{co2} (\num{4} channels $\rightarrow$ \num{12} dimensions including derivatives) and \gls*{ece} (\num{42} channels $\rightarrow$ \num{126} dimensions) lead to an input size of \num{192}. From there, we map to \gls*{ts} with \num{288} dimensions for plasma density and temperature.

\subsection{Spectrogram model development}
\label{subsec:SpectrogramModelDevelopment}

The resulting multi-channel \gls*{ece} spectrograms were used as the input to a \gls*{cnn}, and the multi-channel \gls*{co2} spectrograms were used as the target outputs. We optimized all important hyper-parameters based on the $\mathcal{L}1$ loss to minimize the difference between the ground truth and the estimated outputs on the validation set.

The optimization process of the model involved several key steps:

\begin{itemize}
    \item The model underwent training for up to \num{500} epochs.
    \item We implemented early stopping with a patience threshold of \num{20} epochs, during which we monitored the validation loss for any improvements.
    \item The AdamW optimizer \cite{Loshchilov2018decoupled}, known for decoupling weight decay from the learning rate, was utilized to minimize the $\mathcal{L}1$ loss function.
    \item We conducted a comprehensive hyper-parameter optimization through a randomized search across \num{1000} iterations for all hyper-parameters listed in Table \ref{tab:cnn_hyperparameters}.
\end{itemize}

The exact search space of the hyper-parameters and their optimized values obtained from the randomized search are summarized in Table \ref{tab:cnn_hyperparameters}.

\begin{table}[!htb]
    \renewcommand{\arraystretch}{1.3}
    \centering
    \caption{Optimized hyperparameters for the spectrogram prediction \gls*{cnn} model.}
    \label{tab:cnn_hyperparameters}
    \begin{tabular}{lll}
    \toprule
    Hyper-parameter & Search space & Optimized value \\
    \midrule
    Batch size & \numrange{1}{8}, random integers & \num{2} \\
    Kernel size & \numrange{3}{15} odd integers & \num{7} \\
    Learning rate & \numrange{1e-5}{1} log uniform & \num{0.482e-3} \\
    \midrule
    Final $\mathcal{L}1$ loss & -- & \num{1.2e-3} \\
    \bottomrule
    \end{tabular}
\end{table}

To reduce the amount of training time, we randomly selected \num{518} discharges from the entire dataset to conduct the hyperparameter  optimization. The model with the best performing hyperparameter setting (achieving an $\mathcal{L}1$ loss of \num{1.2e-3} on the validation set) was then re-trained on all available discharges.

The best-performing model is a \gls*{cnn} that transforms the \gls*{ece} spectrograms with \num{40} channels subsequently to \numlist{32;16;8} feature maps and finally to the \gls*{co2} spectrograms with \num{4} channels. For each feature map, 2D filter kernels with a size of $7\times 7$ are used. Batch normalization was used separately for each channel, and parametric $\mathrm{ReLU}$ activation functions were used after each batch normalization layer. The model had in total \num{95823} trainable parameters (i.e., filter kernels for each feature map, batch normalization parameters, and negative slope of the parametric $\mathrm{ReLU}$ activation function). 

%

\subsection{Time-series model development}
\label{subsec:TimeSeriesModelDevelopment}



For the time-series prediction task, we employed a \gls*{mlp} model. The input data to the \gls*{mlp} comprised the \gls*{cer}, \gls*{co2}, \gls*{ece}, \gls*{mse}, and magnetic diagnostics, along with the first and second temporal derivatives of the \gls*{co2} and \gls*{ece} signals, resulting in a total input size of \num{236} dimensions. The target output was the \gls*{ts} diagnostic data, which had \num{80} dimensions representing electron temperature and density across various spatial locations. The target data were augmented by factor \num{2} by using the upper and lower intervals of each sample as additional targets.

The \gls*{mlp} model was trained for a maximum of \num{500} epochs, with an early stopping mechanism implemented to halt the training process if the validation loss did not improve for \num{20} consecutive epochs. The AdamW optimizer \cite{Loshchilov2018decoupled} was employed to minimize the $\mathcal{L}1$ loss function during training. 

As for the spectrogram model, a comprehensive hyperparameter optimization was undertaken using a randomized search approach spanning \num{2000} iterations. The hyper-parameters jointly optimized included the batch size, hidden layer size, dropout rate, and learning rate.

Table \ref{tab:mlp_hyperparameters} summarizes the optimized hyperparameter values obtained from the randomized search process.

\begin{table}[!htb]
    \renewcommand{\arraystretch}{1.3}
    \centering
    \caption{Optimized hyperparameters for the time-series \gls*{mlp} model.}
    \label{tab:mlp_hyperparameters}
    \begin{tabular}{lll}
    \toprule
    Hyper-parameter & Search space & Optimized value \\
    \midrule
    Batch size & \numrange{1}{2048}, powers of 2 & \num{1024} \\
    Hidden layer size & \numrange{192}{2048} integers & \num{952} \\
    Dropout & \numrange{0}{1} uniform & \num{0.076} \\
    Learning rate & \numrange{1e-5}{1} log uniform & \num{1.998e-3} \\
    \midrule
    Final $R^2$ score & -- & \num{0.92} \\
    \bottomrule
    \end{tabular}
\end{table}

\subsection{Uncertainty quantification}
\label{subsec:uq}
To estimate the uncertainty in super-resolution \gls*{ts}, we employed a Bayesian Neural Network (BNN) with the architecture described in the main manuscript. For each \gls*{ts} channel, we calculated the standard deviation of the BNN outputs, particularly focusing on the pedestal area.

\begin{figure}[!htb]
    \centering
    \includegraphics[width=\textwidth]{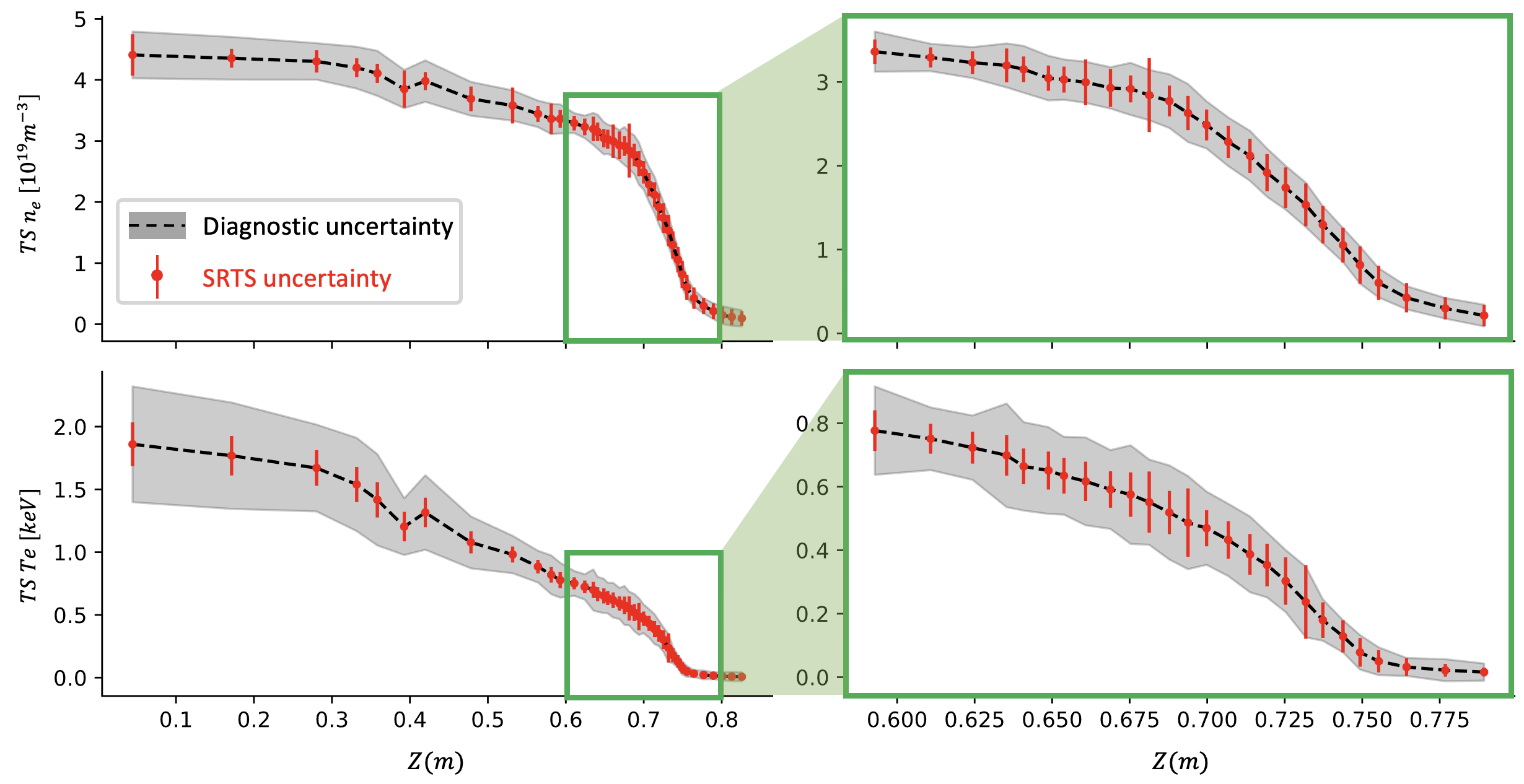}
    \caption{Comparison of the \gls*{srts} and the measured \gls*{ts} diagnostic uncertainty for both electron density $n_e$ and temperature $T_e$. The dashed line shows the average measured \gls*{ts}.}
    \label{fig6}
\end{figure}

Figure \ref{fig6} illustrates the average standard deviations of the neural network outputs per channel over the validation set of discharges, depicted as red error bars. The \gls*{ts} channels are represented by their relative distance from the core of the plasma ($\psi_n=0$) to the edge ($\psi_n=1$). For comparison, we also show the empirically measured diagnostic errors, which include contributions from background light, dark noise, and pulse error.

Although the sources of diagnostic uncertainty differ from those of model uncertainty, our results indicate that the model uncertainty falls in the range of the diagnostic errors that are generally accepted by physicists. This suggests that proposed model provides a reliable estimate of uncertainty that enhances the confidence in the super-resolution \gls*{ts} measurements.

\subsection{Thomson Scattering in ``bunch mode''}
Thomson Scattering, as a popular and reliable diagnostic technique, has successfully measured electron temperatures and electron number densities of plasmas for many years. However, conventional \gls*{ts} techniques operate only at tens of hertz.
To accurately resolve the fast transient dynamics, the Thomson scattering lasers can be fired in a bunch mode, which enabled temporal resolution of up to \SI{10}{\us}. This increase in temporal resolution is achieved by using multiple lasers in the same path with pulses interleaved closely in time. Normally, the lasers are phased to produce pulses at fairly regular intervals (exact regularity is not possible with the specific combination of \SI{20}{\Hz} and \SI{50}{\Hz} lasers being used at \ddd). In bunch mode, the phase shifts are adjusted so that all lasers fire in rapid succession, followed by a cool down. This bunch mode encompasses between 3 and 7 laser pulses depending on the time in the discharges.

\begin{figure}[!htb]
    \centering
    \includegraphics[width=\textwidth]{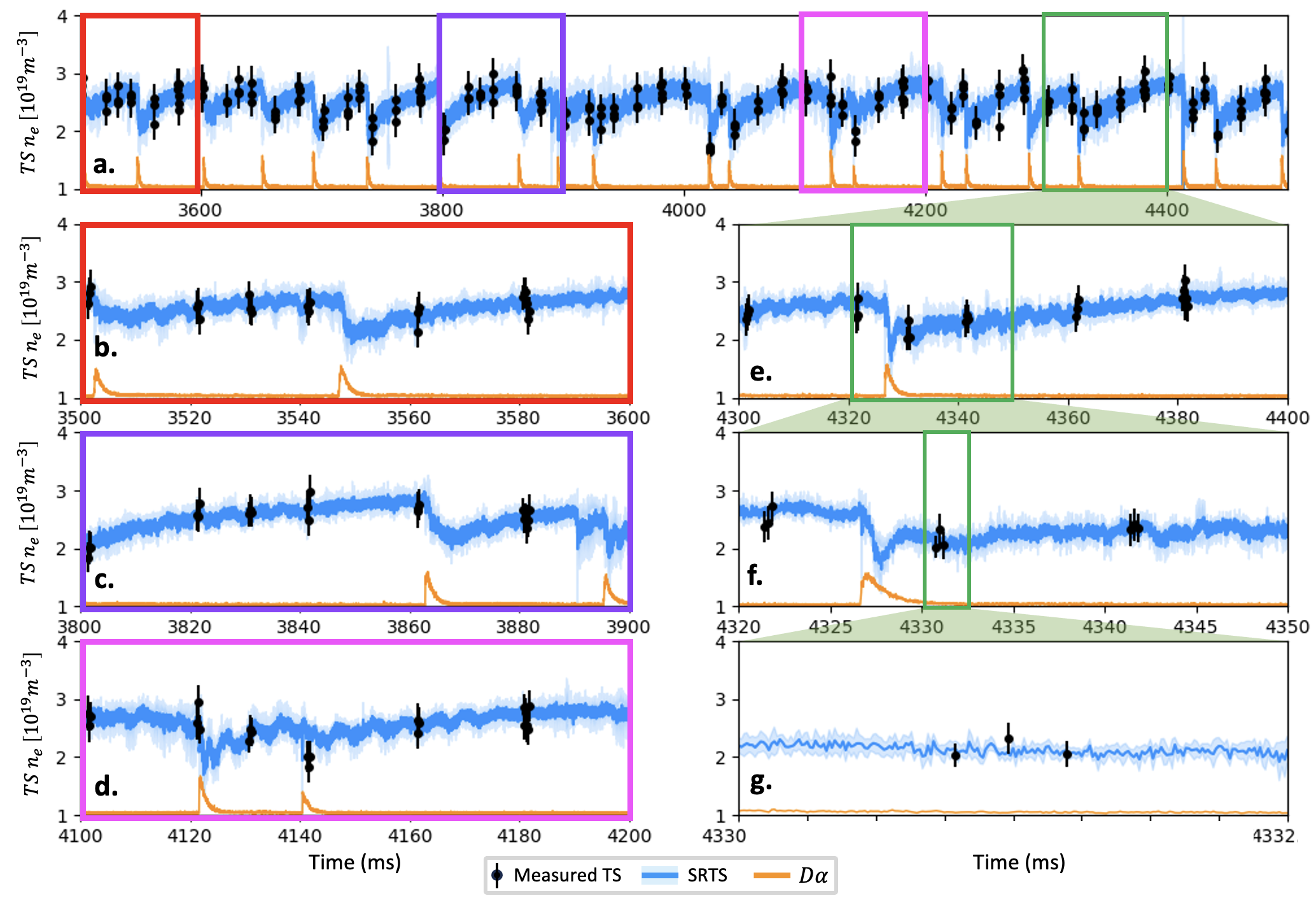}
    \caption{Comparison of the \gls*{srts} and the \gls*{ts} fired in ``bunch mode'' for measuring the electron density in \ddd shot 153761 at the pedestal (Z=0.71m).}
    \label{fig7}
\end{figure}

Figure \ref{fig7}(a) presents the comparison of \gls*{srts} with the measured \gls*{ts} fired at bunch mode for the \ddd shot 153761. Figure \ref{fig7}(b-e) are zoomed in of different window times. Figure \ref{fig7} (f-g) are further zoomed in frames up to one \gls*{ts} fire of the \gls*{ts} lasers. The match between the measured \gls*{ts} and \gls*{srts} confirms the reliability of \gls*{srts} for reconstructing \gls*{ts} when the actual measurement is not available. Also we observe the \gls*{srts} superiority in capturing \gls*{elm}s in comparison to the measured \gls*{ts} even in ``bunch mode''.

\subsection{Research method}    
\label{subsec:method}

To avoid any bias during model development and evaluation, each of the following steps in this research was conducted independently by separate researchers in a feed-forward manner as presented in Figure \ref{fig8}:

\begin{figure}[!htb]
    \centering
    \includegraphics[width=\textwidth]{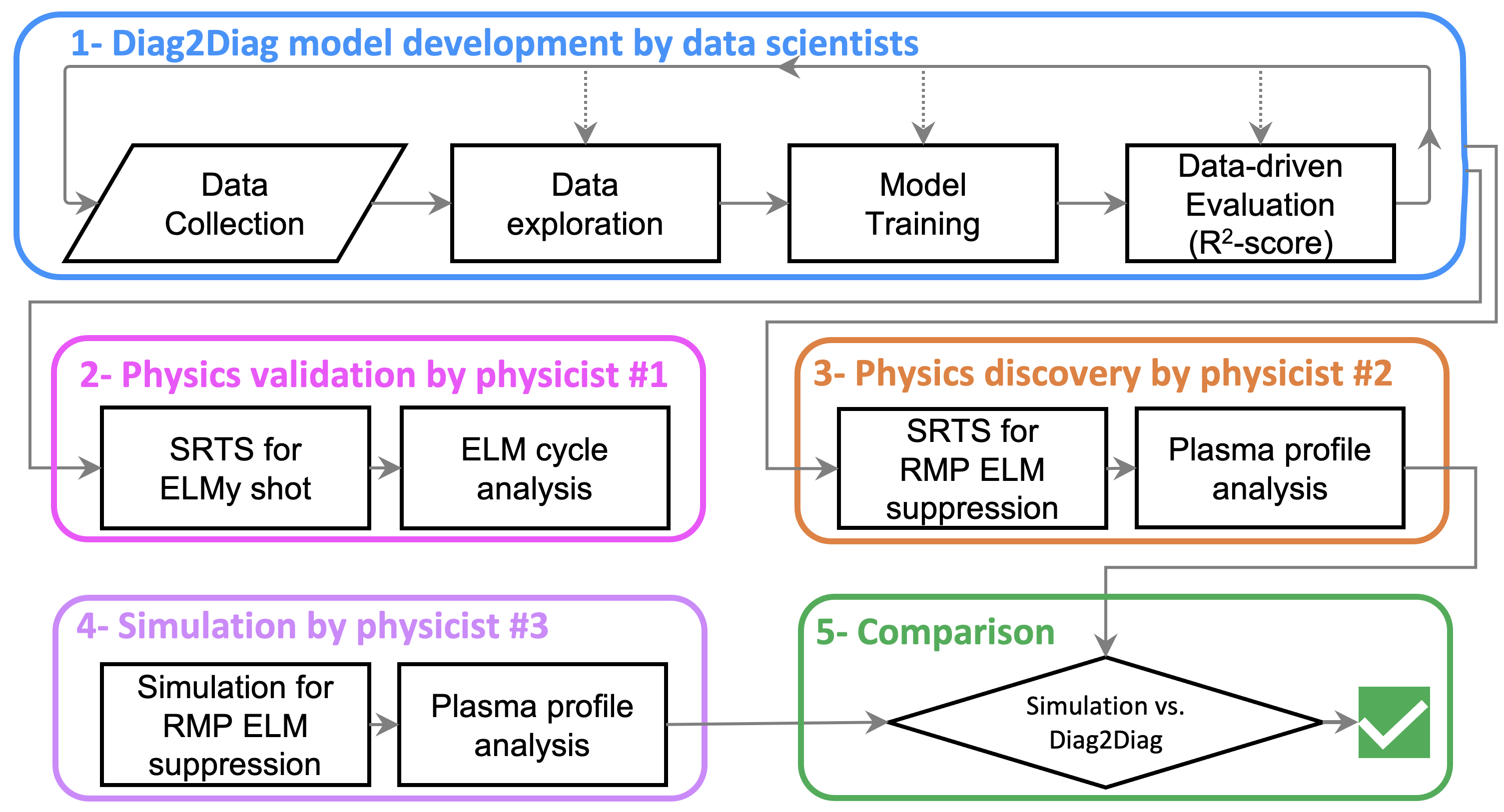}
    \caption{Research steps }
    \label{fig8}
\end{figure}

\begin{enumerate}
    \item The data scientists developed diagnostic dataset for training the neural network aiming for generating synthetic super resolution Thomson Scattering (\gls*{srts}). In this phase, the evaluation metric was simply the similarity between the model's output and the measured TS, whenever the measurement was available.
    \item For physics validation we generated the super resolution diagnostic for a known ELMy discharge and asked an ELM-expert physicist to validate the behavior of the super resolution diagnostic.
    \item The data scientist then delivered the generated super-resolution diagnostic for the target plasma discharge to an experimental physicist to extract the plasma profile from that.
    \item We then asked another physicist with expertise in simulation to obtain the simulation results for the target plasma discharge.
    \item In the final phase, we compared the plasma profiles extracted from our generated diagnostics and the simulation results. They matched nicely!
\end{enumerate}

This indicates that our results are not biased based on prior physics knowledge, and we also did not rework our ML model to match our results with the simulation.

\backmatter





\bmhead{Acknowledgements}

This material is based upon work supported by the U.S. Department of Energy, Office of Science, Office of Fusion Energy Sciences, using the  National Fusion Facility, a DOE Office of Science user facility, under Award DE-FC02-04ER54698. In addition this research was supported by the U.S. Department of Energy, under Awards DE-SC0024527, DE-SC0015480, DE-SC0022270 and DE-SC0022272, as well as the National Research Foundation of Korea (NRF) Award RS-2024-00346024 funded by the Korea government (MSIT).

Disclaimer: This report was prepared as an account of work sponsored by an agency of the United States Government. Neither the United States Government nor any agency thereof, nor any of their employees, makes any warranty, express or implied, or assumes any legal liability or responsibility for the accuracy, completeness, or usefulness of any information, apparatus, product, or process disclosed, or represents that its use would not infringe privately owned rights. Reference herein to any specific commercial product, process, or service by trade name, trademark, manufacturer, or otherwise does not necessarily constitute or imply its endorsement, recommendation, or favoring by the United States Government or any agency thereof. The views and opinions of authors expressed herein do not necessarily state or reflect those of the United States Government or any agency thereof.

\section*{Author contributions}
A.J. is the main author of the manuscript and contributed to developing the multimodal model and data science analyses. S.K. contributed to the physics analysis of Diag2Diag for the RMP mechanism on the plasma boundary. J.S contributed to the general physics analysis and writing the manuscript. Q.H. contributed to the physics simulation of the RMP mechanism on the plasma boundary. M.C. and P.S. contributed to the \ddd data collection and data preprocessing for the multimodal model development and writing the manuscript. A.O.N contributed the physics analysis of Diag2Diag for ELM cycles. Y.S.N and E.K. contributed to the conception of this work, analyses, and writing the manuscript.

\section*{Competing interests}
The authors declare no competing interests.

\section*{Data availability}
The data that support the findings of this study are available from the corresponding
author upon reasonable request. The source code for the models developed in this work can be found in \cite{diag2diag_github}.

\bibliography{sn-bibliography}


\end{document}